\documentclass[11pt,journal,draftcls,onecolumn,peerreviewca]{IEEEtran}
\usepackage{amsfonts}
\usepackage{amsmath}
\usepackage{amssymb}
\usepackage{algorithm}
\usepackage{algorithmic}
\usepackage{bm}
\usepackage{graphicx}
\usepackage{color, soul}
\usepackage{stfloats}
\usepackage[numbers,sort&compress]{natbib}
\usepackage[amsmath,thmmarks]{ntheorem}
\usepackage{theorem}
\usepackage{subfigure}
\usepackage[justification=centering]{caption}

\theoremheaderfont{\sc}\theorembodyfont{\upshape}
\theoremstyle{nonumberplain}
\theoremseparator{}
\theoremsymbol{\rule{1ex}{1ex}}

\begin{document}
\title{A Two-stage Approach to Estimate CFO and Channel with One-bit ADCs}

\author{Jiang Zhu, Hangting Cao and Zhiwei Xu
\thanks{}}


\maketitle

\begin{abstract}
In this letter, we propose a two-stage approach to estimate the carrier frequency offset (CFO) and channel with one-bit analog-to-digital converters (ADCs). Firstly, a simple metric which is only a function of the CFO is proposed, and the CFO is estimated via solving the one-dimensional optimization problem. Secondly, the generalized approximate message passing (GAMP) algorithm combined with expectation maximization (EM) method is utilized to estimate the channel. In order to provide a benchmark of our proposed algorithm in terms of the CFO estimation, the corresponding Cram\'er-Rao bound (CRB) is derived. Furthermore, numerical results demonstrate the effectiveness of the proposed approach when applied to the general Gaussian channel and mmWave channel.
\end{abstract}

{\bf{keywords}}: CFO, channel estimation, millimeter wave system, one-bit quantization

\section{Introduction}
To provide a high-speed data rate in celluar systems, the mmWave multiple input multiple output (MIMO) system has been proposed as the key technology of the fifth generation (5G) cellular system \cite{Zhang, Fang}. Because of the larger bandwidths that accompany mmWave, the cost and power consumption are huge due to high precision (e.g., 10-12 bits) analog-to-digital converters (ADCs) \cite{Rangan1}. As a result, a low precision (e.g., 1-4 bits) ADC is employed to relieve this ADC bottleneck \cite{Singh1, Singh2}. However, as low precision quantization is severely nonlinear, traditional algorithms designed for high precision systems can not be applied directly because of significant performance degradation. As a consequence, new signal processing algorithms dealing with channel estimation and transmit precoding have been proposed, which work well in systems with low precision ADCs \cite{Risi, Li, Mollen, Fangqing1}. For the channel estimation in mmWave systems, it can be regarded as one-bit compressed sensing (CS) problems \cite{Boyd, Boufounos, Fangone1, Fangone2, Fangone3}, as the mmWave MIMO channel is approximately sparse in angle domain \cite{Heath1}. Therefore, many CS-based algorithms have been proposed to estimate the mmWave MIMO channel. In \cite{Mo, Heath}, a modified expectation maximization (EM) algorithm and approximate message passing (AMP) algorithms are utilized to solve the channel estimation problem in mmWave MIMO systems.

In practice, the carrier frequencies between the local oscillators at the TX and the RX can be mismatched, which results in carrier frequency offset (CFO) impairing the phase of the channel measurements in systems. One approach to dealing with the above problem is to correct the CFO before channel estimation, which is impractical because the mmWave systems always work at low SNR prior to channel estimation \cite{Myers1}. As a result, several works have studied the joint CFO and channel estimation \cite{Myers1, Myers, Javier, Pajovic}. In \cite{Myers}, a generalized AMP (GAMP) algorithm to jointly estimate the CFO and channel in mmWave narrowband systems with one-bit ADCs is developed. It utilizes a lifting technique which increases the problem's dimension. In \cite{Myers1}, an algorithm called PBiGAMP is proposed to jointly estimate CFO and wideband channel, which has a much lower computational complexity.


In this letter, we propose a two-stage approach to estimate the CFO and channel with one-bit ADCs. Firstly, we utilize Bussgang decomposition theorem which transforms the non-linear model into a linear model \cite{Andrew}, and the CFO is estimated via solving the one-dimensional optimization problem. Secondly, by fixing the CFO with the estimated CFO, we apply the GAMP-EM algorithm \cite{Rangan,Schniter} to estimate the channel. Besides, the CRB is also derived for evaluating the performance of our algorithm in terms of CFO estimation. One appealing advantage of the proposed method is that both the CFO and channel can be estimated accurately without increasing the problem's dimension. Numerical results show the effectiveness of the proposed two-stage approach, i.e., the estimation performance degradation in terms of the CFO and channel of the proposed method is marginal, compared to the benchmarks such as the CRB and the CFO-known (oracle) algorithm.
\section{Algorithm}
In this section, the problem model and algorithm are introduced. Consider a $N_t\times N_r$ MIMO system with one-bit ADCs and let $\omega_e$ denote the CFO. For a training block $\mathbf{T}\in {\mathbb C}^{N_{t}\times N_{p}}$, the observation $\mathbf{Y}\in {\mathbb C}^{N_{r}\times N_p}$ obtained at ADCs is
\begin{align}\label{model}
\mathbf Y = {\rm csgn}(\mathbf H \mathbf T {\rm diag}(a_{N_p}(\omega_e))+\mathbf W),
\end{align}
where $\mathbf H \in {\mathbb C}^{N_r\times N_t}$ is the channel matrix, ${\rm csgn}(\cdot)$ is an element-wise quantization function given by ${\rm csgn}(x)={\rm sgn}({\rm Re}\{x\}) + {\rm j}{\rm sgn}({\rm Im}\{x\})$ with ${\rm sgn}(\cdot)$ being the signum function, $a_{N}(\theta)$ is the Vandermonde vector given by $a_{N}(\theta) = [1, {\rm e}^{{\rm j}\theta}, {\rm e}^{{\rm j}2\theta}, \dots , {\rm e}^{{\rm j}(N-1)\theta}]^{\rm T}$ and $\mathbf W$ is the additive white Gaussian noise, i.e., ${W}_{ij}\sim \mathcal{CN}(0,2\sigma_w^2)$ with $\sigma_w^2$ being known. We aim to estimate the CFO $\omega_e$ and channel $\mathbf H$ based on the observation $\mathbf Y$ and the training block $\mathbf T$.

At the beginning, we reformulate model (\ref{model}) to a real-valued form. Utilizing the property ${\rm vec}(\mathbf A \mathbf B \mathbf C)=({\mathbf C}^{\rm T} \otimes \mathbf A){\rm vec}(\mathbf B)$, it can be transformed to a vector form firstly as
\begin{align}\label{model_vector}
\mathbf{y}_v={\rm csgn}(\mathbf{F}\mathbf{h}_v+\mathbf{w}_v),
\end{align}
where $\mathbf{y}_v = \rm{vec}(\mathbf{Y})$, $\mathbf{F} = \mathbf{B}^{\rm{T}} \otimes \mathbf{I}_{N_{r}}$, $\mathbf{B} = \mathbf{T}{\rm{diag}}(\mathbf{a}_{N_p}(\omega_e))$, $\mathbf{h}_v = \rm{vec}(\mathbf{H})$ and $\mathbf{w}_v = \rm{vec}(\mathbf{W})$. By defining
\begin{subequations}       
\begin{align}
&\mathbf{y}=\left[                 
  \begin{array}{c}   
    \mathbf{y}_{v}^{\rm R}\\  
    \mathbf{y}_{v}^{\rm I}\\  
  \end{array}
\right],
&&\mathbf{h}=\left[
\begin{array}{c}   
    \mathbf{h}_{v}^{\rm R}\\  
    \mathbf{h}_{v}^{\rm I}\\  
  \end{array}
\right],      \\              
&\mathbf{w}=\left[
\begin{array}{c}   
    \mathbf{w}_{v}^{\rm R}\\  
    \mathbf{w}_{v}^{\rm I}\\  
  \end{array}
\right],
&&\mathbf{D}=\left[
\begin{array}{cc}   
    \mathbf{F}_{\rm R} & -\mathbf{F}_{\rm I}\\  
    \mathbf{F}_{\rm I} & \mathbf{F}_{\rm R} \\  
  \end{array}
\right],
\end{align}
\end{subequations}
a real-valued equivalent model
\begin{align}\label{realmodel}
\mathbf{y}={\rm sgn}(\mathbf{D}\mathbf{h}+\mathbf{w}),
\end{align}
is obtained, where $\mathbf{w}\sim {\mathcal N}(0,{ \mathbf{C}}_w)$ and $\mathbf{C}_w = \sigma_w^2\mathbf I$.
\subsection{CFO Estimation}
Before performing the channel estimation, we estimate the CFO first. We assume that the prior distribution of ${\mathbf h}$ follows ${\mathbf h}\sim {\mathcal N}({\mathbf 0},{\mathbf {C}_h})$ and we use the method proposed in \cite{Andrew} to linearize the model as
\begin{align}\label{linear-model}
\mathbf y = \mathbf G \mathbf h + \mathbf e,
\end{align}
where $\mathbf G$ is the linearization matrix and $\mathbf e$ is a residual error vector consisting of noise and linearization artifacts. According to \cite{Andrew}, $\mathbf G$ is calculated as
\begin{align}
\mathbf{G}=(\frac{2}{\pi})^{1/2}{\rm diag}\left(({\rm diag}(\mathbf{C}_z))^{-1/2}\right)\mathbf{D},\label{G}
\end{align}
where $\mathbf{C}_z=\mathbf{D}\mathbf{C}_h\mathbf{D}^{\rm T}+\mathbf{C}_w$.
To estimate the CFO, we maximize the expected energy (taken with respect to $\mathbf h$) of the output of the matched filtering of the observation $\mathbf y$, which can be expressed as
\begin{align}\label{iniproblem}
\max\limits_{\omega_e} {\mathbb E}_{\mathbf h}[{\Vert{\mathbf y}^{\rm T}\mathbf G \mathbf h\Vert}_2^2].
\end{align}
Assuming $\mathbf C_h = \sigma_h^2\mathbf I$ and omitting the constant coefficient, (\ref{iniproblem}) can be simplified as
\begin{align}\label{optimization}
\max\limits_{\omega_e}{\Vert{\mathbf G}^{\rm T}\mathbf y\Vert}_2^2.
\end{align}
Furthermore, for an independent and identically distributed (iid) QPSK training block $\mathbf T$, the optimization problem (\ref{optimization}) can be simplified further. First, we rewrite $\mathbf D$ as
\begin{equation}
\mathbf{D}=\left[
\begin{array}{cc}   
    \mathbf{B}_{\rm R}^{\rm T} & -\mathbf{B}_{\rm I}^{\rm T}\\  
    \mathbf{B}_{\rm I}^{\rm T} & \mathbf{B}_{\rm R}^{\rm T} \\  
  \end{array}
\right]\otimes {\mathbf I}_{N_{r}}.
\end{equation}
From equation (\ref{G}), we extract the diagonal elements of $\mathbf C_z$ and using the property $(\mathbf A\otimes \mathbf B)(\mathbf C \otimes \mathbf D)= (\mathbf A \mathbf C)\otimes(\mathbf B \mathbf D)$, we obtain
\begin{equation}
\mathbf D {\mathbf D}^{\rm T} = \left[
\begin{array}{cc}
    \mathbf{B}_{\rm R}^{\rm T}\mathbf{B}_{\rm R}+ \mathbf{B}_{\rm I}^{\rm T}\mathbf{B}_{\rm I} & \mathbf{B}_{\rm R}^{\rm T}\mathbf{B}_{\rm I}-\mathbf{B}_{\rm I}^{\rm T}\mathbf{B}_{\rm R}\\  
    \mathbf{B}_{\rm I}^{\rm T}\mathbf{B}_{\rm R}-\mathbf{B}_{\rm R}^{\rm T}\mathbf{B}_{\rm I}&\mathbf{B}_{\rm R}^{\rm T}\mathbf{B}_{\rm R}+ \mathbf{B}_{\rm I}^{\rm T}\mathbf{B}_{\rm I} \\  
  \end{array}
\right]\otimes{\mathbf I}_{N_{r}}.\notag
\end{equation}
Recall that
\begin{align}
\mathbf{B}= \mathbf{T}{\rm{diag}}(\mathbf{a}_{N_p}(\omega_e))=[\mathbf t_1, {\rm e}^{{\rm j}\omega_e}{\mathbf t}_2, \dots, {\rm e}^{{\rm j}(N_p-1)\omega_e}{\mathbf t}_{N_p}],\notag
\end{align}
where ${\mathbf t}_i$ denotes the $i$th column of $\mathbf T$. The diagonal elements of ${\mathbf C}_z$ are
\begin{equation}\label{C_z}
{\rm diag}({\mathbf C}_z) = \sigma_h^2\left[
\begin{array}{cc}
    {\mathbf c}_t \\  
    {\mathbf c}_t\\  
  \end{array}
\right]\otimes{\mathbf 1}_{N_{r}}+{\rm diag}({\mathbf C}_w),
\end{equation}
where
\begin{align}
{\mathbf c}_t &= [{\Vert\mathbf t_1\Vert}_2^2, {\Vert e^{j\omega_e}\mathbf t_2\Vert}_2^2, \dots, {\Vert e^{j(N_p-1)\omega_e}\mathbf t_{N_p}\Vert}_2^2]^{\rm T},\notag\\
&=[{\Vert\mathbf t_1\Vert}_2^2, {\Vert\mathbf t_2\Vert}_2^2, \dots, {\Vert\mathbf t_{N_p}\Vert}_2^2]^{\rm T}.\label{Ct}
\end{align}
For an iid QPSK training block $\mathbf T$ which takes values in $\{\pm1\pm{\rm j}\}$, ${\Vert{\mathbf t}_i\Vert}_2^2$ is equal to $2N_{t}$. Therefore, ${\rm diag}({\mathbf C}_z)$ is simplified as ${\rm diag}({\mathbf C}_z) = (2\sigma_h^2N_{t}+\sigma_w^2)\textbf{1}$,
and $\mathbf G$ is simplified as $\mathbf G = (\frac{2}{\pi})^{1/2}(2\sigma_h^2N_{t}+\sigma_w^2)^{-1/2}\mathbf D$.
As a result, the optimization problem (\ref{optimization}) is further simplified as
\begin{align}\label{finalopt}
\max\limits_{\omega_e}{\Vert{\mathbf D}^{\rm T}\mathbf y\Vert}^2_2.
\end{align}
To solve the problem (\ref{optimization}) or (\ref{finalopt}), we adopt two steps \cite{Mamandipoor, Han}: Detection and Refinement.

\emph{Detection}: The Detection step includes coarse detection and refined detection. Firstly, we solve the optimization problem (\ref{optimization}) or (\ref{finalopt}) and obtain a coarse estimate ${\hat \omega}_c$ of $\omega_e$ by restricting it to a discrete set denoted by $\{0,\frac{2\pi}{N_1},\dots,\frac{2\pi(N_1-1)}{N_1}\}$. Secondly, we implement a refined detection over the frequencies around ${\hat \omega}_c$. We solve the same problem again, but restrict $\omega_e$ to the discrete set $\{{\hat \omega}_c-\frac{(N_2-1)2\pi}{N_1N_2},{\hat \omega}_c-\frac{(N_2-2)2\pi}{N_1N_2},\dots,{\hat \omega}_c+\frac{(N_2-1)2\pi}{N_1N_2}\}$ this time, and finally update ${\hat \omega}_c$ as ${\hat \omega}_r$. We found that $N_1=300$ and $N_2=10$ work well for a large number of problems. Due to page limitations, we refer interested readers to the supplementary materials for more details about the parameters $N_1$ and $N_2$.

\emph{Refinement}: Numerical results show that problem (\ref{optimization}) or (\ref{finalopt}) is locally concave around the global optimum. As a result, the estimate ${\hat \omega}_r$ given by the Detection step is used as an initial point and the gradient descent algortihm is performed to refine the estimate ${\hat \omega}_r$ as ${\hat \omega}_e$.

Furthermore, in order to evaluate the performance of the proposed approach for CFO estimation, the CRB of CFO ${\rm CRB}(\omega_e)$ (\ref{CRB}) is derived in Section \ref{sectionCRB}.
\subsection{Channel Estimation}
In this section, we transform the channel estimation problem to a general model
\begin{align}\label{generalmodel}
\mathbf y={\rm sgn}(\mathbf A \mathbf x+\mathbf w+\boldsymbol \tau),
\end{align}
and then apply the GAMP-EM algorithm directly \cite{Schniter,cht}, where the EM method recovers the nuisance parameters of the prior distribution of $\mathbf x$. In our channel estimation problem, we set $\boldsymbol \tau=\textbf{0}$ and consider two kinds of channel: The general Gaussian channel and the mmWave channel. More details are provided below.
\subsubsection{General Gaussian Channel}
For the general Gaussian channel, the channel matrix $\mathbf H$ follows a zero-mean Gaussian distribution, i.e., ${H}_{ij} \sim {\mathcal {CN}}(0,2\sigma_h^2)$ with $\sigma_h^2$ being unknown. We apply the GAMP-EM algorithm directly on $\mathbf h$. Therefore, the corresponding $\mathbf A$ and $\mathbf x$ in model (\ref{generalmodel}) are $\mathbf D$ and $\mathbf h$ in the model (\ref{realmodel}), respectively. For the denoising step in the GAMP algorithm, we denoise the noisy signal with the prior of ${\mathbf h}$ being Gaussian.
\subsubsection{mmWave Channel}
A narrowband mmWave channel can be modeled by a ray-based model \cite{Myers}. For a propagation environment having $N_c$ clusters and $K_n$ rays in the $n$th cluster, the channel matrix $\mathbf H$ is described as
\begin{subequations}
\begin{align}
&\mathbf H = \frac{1}{\sqrt{N_c}}\sum\limits_{n=1}^{N_c}\frac{1}{\sqrt{K_n}}\sum\limits_{m=1}^{K_n}\gamma_{n,m}{\mathbf a}_{N_r}(\omega_{r,m,n}){\mathbf a}_{N_t}^{\rm H}(\omega_{t,m,n}),\notag\\
&\omega_{r,n,m} = \frac{2\pi d}{\lambda}{\rm sin}(\theta_{r,n,m}),\quad
\omega_{t,n,m} = \frac{2\pi d}{\lambda}{\rm sin}(\theta_{t,n,m}).\notag
\end{align}
\end{subequations}
Here, $\gamma_{n,m}$, $\theta_{r,m,n}$ and $\theta_{t,m,n}$ are the complex gain, angle-of-arrival and angle-of-departure of the $m$th ray in the $n$th cluster, respectively. $\lambda$ and $d$ denote the carrier wavelength and antenna spacing.

For the mmWave MIMO channel, its beamspace representation of channel matrix $\mathbf H$ is
\begin{align}\label{beam-space}
\mathbf H = {\mathbf U}_{N_{r}} \mathbf C {\mathbf U}_{N_{t}}^{\rm H},
\end{align}
where ${\mathbf U}_{N_{r}}\in {\mathbb C}^{N_r\times N_r}$ and ${\mathbf U}_{N_{t}}\in {\mathbb C}^{N_t\times N_t}$ are unitary Discrete Fourier Transform matrices. Since the mmWave MIMO channel is approximately sparse in angle domain, ${\mathbf C}\in {\mathbb C}^{N_r\times N_t}$ in (\ref{beam-space}) is a sparse matrix. We assume that ${C}_{ij}$ follows the Bernoulli-Gaussian distribution and apply the GAMP-EM algorithm on $\mathbf c$ instead of $\mathbf h$, where $\mathbf c=[({\mathbf c}_v^{\rm R})^{\rm T},({\mathbf c}_v^{\rm I})^{\rm T}]^{\rm T}$ and ${\mathbf c}_v={\rm vec}(\mathbf C)$. In this case, we can obtain a model ${\mathbf y}_v={\rm csgn}({\mathbf F}_c{\mathbf c}_v+{\mathbf w}_v)$ similar to model (\ref{model_vector}), where ${\mathbf F}_c = {\mathbf B}_c^{\rm T}\otimes {\mathbf U}_{N_r}$ and ${\mathbf B}_c = {\mathbf U}_{N_t}^{\rm H}\mathbf{T}{\rm{diag}}(\mathbf{a}_{N_p}(\omega_e))$. Then following the similar steps through transforming  model (\ref{model_vector}) to (\ref{realmodel}), a real-valued equivalent model $\mathbf y = {\rm sgn}({\mathbf D}_c\mathbf c+\mathbf w)$ is obtained. Therefore, for the GAMP-EM algorithm, the corresponding $\mathbf A$ and $\mathbf x$ in model (\ref{generalmodel}) are ${\mathbf D}_c$ and $\mathbf c$, respectively.
\section{Cram\'er-Rao bound}\label{sectionCRB}
In this part, the details about the calculation of the CRB of CFO are presented. First we start from the problem model (\ref{realmodel}) and to be more concretely, the two parts ${\mathbf F}_{\rm R}$ and ${\mathbf F}_{\rm I}$ of matrix $\mathbf D$ are
\begin{align}
\mathbf{F}_{\rm R}&=\mathbf{B}_{\rm R}^{\rm T}\otimes \mathbf{I}_{N_{r}}=\left({\rm diag}(\bar{\mathbf c})\mathbf{T}_{\rm R}^{\rm T}-{\rm diag}(\bar{\mathbf s})\mathbf{T}_{\rm I}^{\rm T}\right)\otimes \mathbf{I}_{N_{r}},\notag
\end{align}
\begin{align}
\mathbf{F}_{\rm I}&=\mathbf{B}_{\rm I}^{\rm T}\otimes \mathbf{I}_{N_{r}}=\left({\rm diag}(\bar{\mathbf s})\mathbf{T}_{\rm R}^{\rm T}+{\rm diag}(\bar{\mathbf c})\mathbf{T}_{\rm I}^{\rm T}\right)\otimes \mathbf{I}_{N_{r}},\notag
\end{align}
where the $i$th element of $\bar{\mathbf{s}}$ and $\bar{\mathbf{c}}$ are ${\bar{s}}_i={\rm sin}(i-1)\omega_e$ and ${\bar{c}}_i={\rm cos}(i-1)\omega_e$, for $i=1,2,\dots,N_p$.
Let $\mathbf{d}_i^{\rm T}$ denote the $i$th row of $\mathbf{D}$, the likelihood function $\rm{Pr}(\mathbf{y};\mathbf{h},\omega_e)$ is ${\rm{Pr}}(\mathbf{y};\mathbf{h},\omega_e)=\prod\limits_{i}\Phi(y_i\frac{\mathbf{d}_i^{\rm T}\mathbf{h}}{\sigma_w})$
and the corresponding log-likelihood function $l(\mathbf{y};\mathbf{h},\omega_e)$ is given by
\begin{align}
l(\mathbf{y};\mathbf{h},\omega_e) = \sum\limits_{i}{\rm{log}}\Phi(y_i\frac{\mathbf{d}_i^{\rm T}\mathbf{h}}{\sigma_w}).
\end{align}
By defining $\phi_i=\frac{1}{2\pi\sigma_w^2}(\frac{1}{\Phi(\frac{\mathbf{d}_i^{\rm T}\mathbf{h}}{\sigma_w})}+\frac{1}{\Phi(-\frac{\mathbf{d}_i^{\rm T}\mathbf{h}}{\sigma_w})})e^{-\frac{(\mathbf{d}_i^{\rm T}\mathbf{h})^2}{\sigma_w^2}}$, we calculate
\begin{subequations}
\begin{align}
&{\rm{E}}_{\mathbf{y}}[\nabla^2_{\mathbf{h}}l(\mathbf{y};\mathbf{h},\omega_e)]=-\sum\limits_{i}\phi_i\mathbf{d}_i\mathbf{d}_i^{\rm{T}},\\
&{\rm{E}}_{\mathbf{y}}[\nabla^2_{\omega_e}l(\mathbf{y};\mathbf{h},\omega_e)]=-\sum\limits_{i}\phi_i(\dot {\mathbf{d}}_i^{\rm T}\mathbf{h})^2,\\
&{\rm{E}}_{\mathbf{y}}[\nabla^2_{\omega_e\mathbf{h}}l(\mathbf{y};\mathbf{h},\omega_e)]=-\sum\limits_{i}\phi_i\dot {\mathbf{d}}_i^{\rm T}\mathbf{h}\mathbf{d}_i,
\end{align}
\end{subequations}
where $\dot{\mathbf{d}}_i^{\rm T}$ is the $i$th row of $\dot{\mathbf{D}}$ and $\dot {\mathbf{D}}$ is
\begin{align}\label{patial}
\dot {\mathbf{D}}=\frac{\partial \mathbf{D}}{\partial \omega_e}=\left[
\begin{array}{cc}
\frac{\partial \mathbf{F}_{\rm R}}{\partial \omega_e} & -\frac{\partial \mathbf{F}_{\rm I}}{\partial \omega_e}\\
\frac{\partial \mathbf{F}_{\rm I}}{\partial \omega_e} & \frac{\partial \mathbf{F}_{\rm R}}{\partial \omega_e}
\end{array}
\right].
\end{align}
Here,
\begin{align}
\frac{\partial \mathbf{F}_{\rm R}}{\partial \omega_e}={\rm diag}(\mathbf a)\left({-\rm diag}(\bar{\mathbf s})\mathbf{T}_{\rm R}^{\rm T}-{\rm diag}(\bar{\mathbf c})\mathbf{T}_{\rm I}^{\rm T}\right)\otimes \mathbf{I}_{N_{r}},\notag
\end{align}
\begin{align}
\frac{\partial \mathbf{F}_{\rm I}}{\partial \omega_e}={\rm diag}(\mathbf a)\left({\rm diag}(\bar{\mathbf c})\mathbf{T}_{\rm R}^{\rm T}-{\rm diag}(\bar{\mathbf s})\mathbf{T}_{\rm I}^{\rm T}\right)\otimes \mathbf{I}_{N_{r}},\notag
\end{align}
where the $i$th element of $\mathbf a$ is ${a}_i=i-1$ for $i=1,2,\dots,N_p$.
Let $\mathbf{z}=[\omega_e,\mathbf{h}^{\rm T}]^{\rm T}$, the Fisher Information Matrix (FIM) is
\begin{align}
\mathbf{J}(\mathbf{z})=-{\rm{E}}_{\mathbf{y}}[\nabla^2_{\mathbf{z}}l(\mathbf{y};\mathbf{z})]=\mathbf{M}^{\rm{T}} {\boldsymbol \Lambda} \mathbf{M},
\end{align}
where
\begin{align}
\mathbf{M}=[\dot{\mathbf{D}}\mathbf{h},\mathbf{D}]=
\left[
\begin{array}{ccc}
\frac{\partial \mathbf{F}_{\rm R}}{\partial \omega_e}\mathbf{h}_v^{\rm R}-\frac{\partial \mathbf{F}_{\rm I}}{\partial \omega_e}\mathbf{h}_v^{\rm I}&\mathbf{F}_{\rm R} & -\mathbf{F}_{\rm I}\\
\frac{\partial \mathbf{F}_{\rm I}}{\partial \omega_e}\mathbf{h}_v^{\rm R}+\frac{\partial \mathbf{F}_{\rm R}}{\partial \omega_e}\mathbf{h}_v^{\rm I}&\mathbf{F}_{\rm I} & \mathbf{F}_{\rm R}
\end{array}
\right],\notag
\end{align}
and $\Lambda$ is a diagonal matrix with elements ${\Lambda}_{ii} = \phi_i$.
The CRB is equal to ${\mathbf J}^{-1}$. For the CFO, utilizing a well-known matrix inversion relation \cite{Wiesel}, its CRB is
\begin{align}\label{CRB}
{\rm CRB}(\omega_e) = ({J}_{\omega_e\omega_e}-{\mathbf J}_{\omega_e\mathbf h}{\mathbf J}_{\mathbf h\mathbf h}^{-1}{\mathbf J}_{\mathbf h\omega_e})^{-1},
\end{align}
where
\begin{subequations}
\begin{align}
&{J}_{\omega_e\omega_e}={\mathbf h}^{\rm T}\dot{\mathbf D}^{\rm T}\Lambda\dot{\mathbf D}{\mathbf h},\quad {\mathbf J}_{\omega_e\mathbf h}={\mathbf h}^{\rm T}\dot{\mathbf D}^{\rm T}\Lambda{\mathbf D},\notag\\
&{\mathbf J}_{\mathbf h\omega_e}={\mathbf D}^{\rm T}\Lambda\dot{\mathbf D}{\mathbf h},\quad {\mathbf J}_{\mathbf h\mathbf h}={\mathbf D}^{\rm T}\Lambda{\mathbf D}.\notag
\end{align}
\end{subequations}
\section{Numerical Simulation}\label{NS}
In this section, the performance of the proposed algorithm is evaluated by applying to both the general Gaussian channel and mmWave channel, which is measured by the mean square error (MSE) of CFO estimate $\hat{\omega}_e$ and the normalized MSE (NMSE) of channel estimate $\hat{\mathbf H}$($\hat{\mathbf C}$). Meanwhile, the CRB of CFO (\ref{CRB}) is also plotted. In our simulations, the MSE of CFO estimate is denoted by ${\rm MSE}(\omega_e)=\mathbb E \left[(\omega_e-{\hat \omega_e})^2\right]$ and the NMSE of channel estimate is denoted by ${\rm NMSE}(\mathbf H)=\mathbb E[{\Vert\mathbf H -\hat{\mathbf H} \Vert}_{\rm F}^2/{\Vert \mathbf H\Vert}_F^2]$ for the Gaussian channel and ${\rm NMSE}(\mathbf C)=\mathbb E[{\Vert\mathbf C -\hat{\mathbf C} \Vert}_{\rm F}^2/{\Vert \mathbf C\Vert}_F^2]$ for the mmWave channel. An iid QPSK training block $\mathbf T$ is used in our experiments and the system parameters are set as follows: $N_r=16$, $N_t=16$, $\sigma_h^2=0.5$. We choose $\omega_e = 2\pi\times 4.15\%$, where the percentage we choose is to make $\omega_e$ maximally off the grids in Detection step. We set $\sigma_w^2$ such that ${\rm SNR} = 10{\rm log}_{10}\frac{{\Vert \mathbf D \mathbf h\Vert}_2^2}{2N_r N_p \sigma_w^2}$, where ${\rm SNR}=\{0,5,10\} {\rm dB}$. All the results are averaged over $500$ Monte Carlo (MC) trials.
\subsection{General Gaussian Channel}
\begin{figure}
\centering{\includegraphics[width=2.5in]{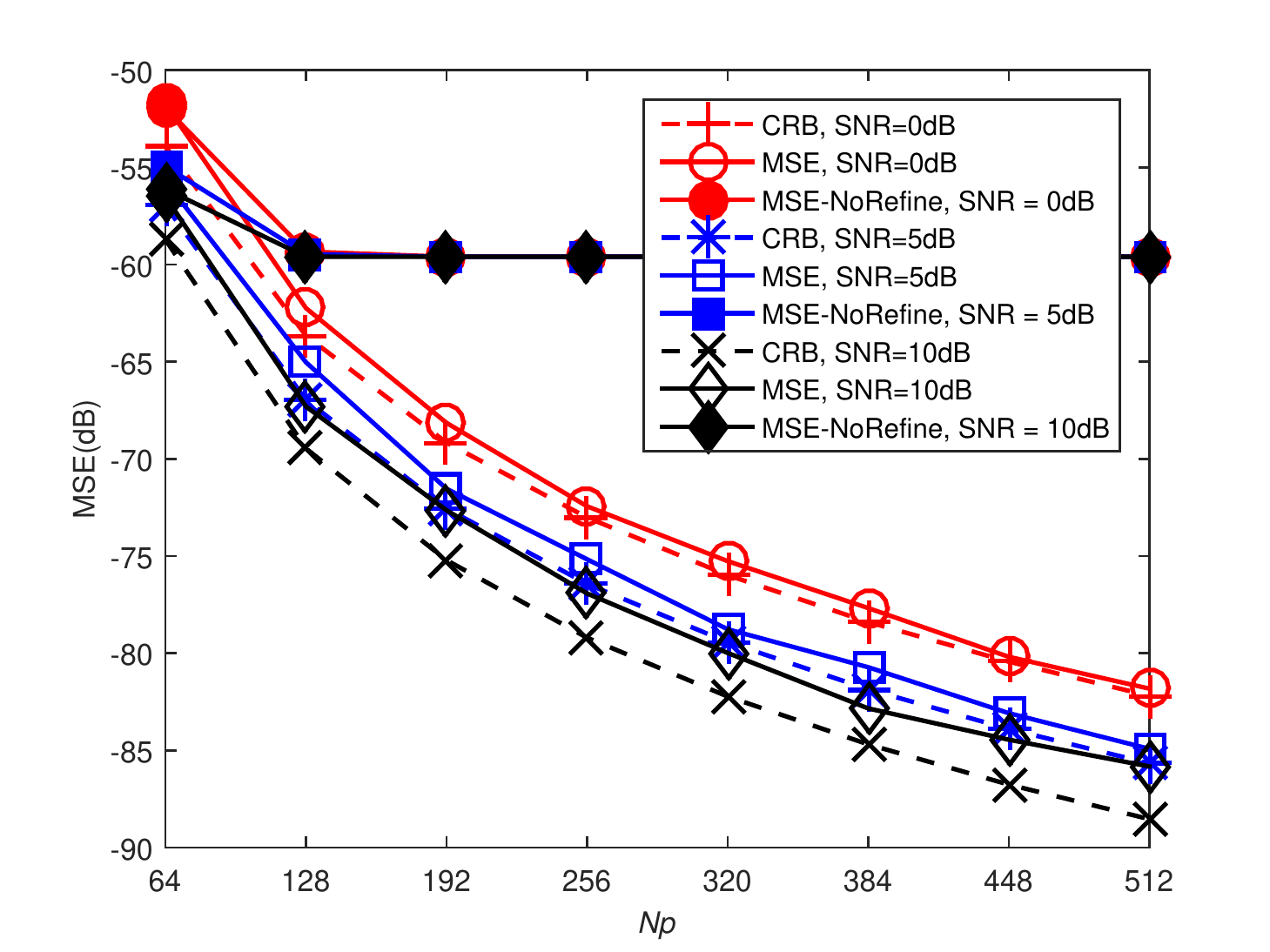}}
\caption{The MSE and CRB of the CFO versus the length of training block $N_p$ with different SNRs for the general Gaussian channel.
}\label{MSEw_Gaussian}
\end{figure}

\begin{figure}
\centering{\includegraphics[width=2.5in]{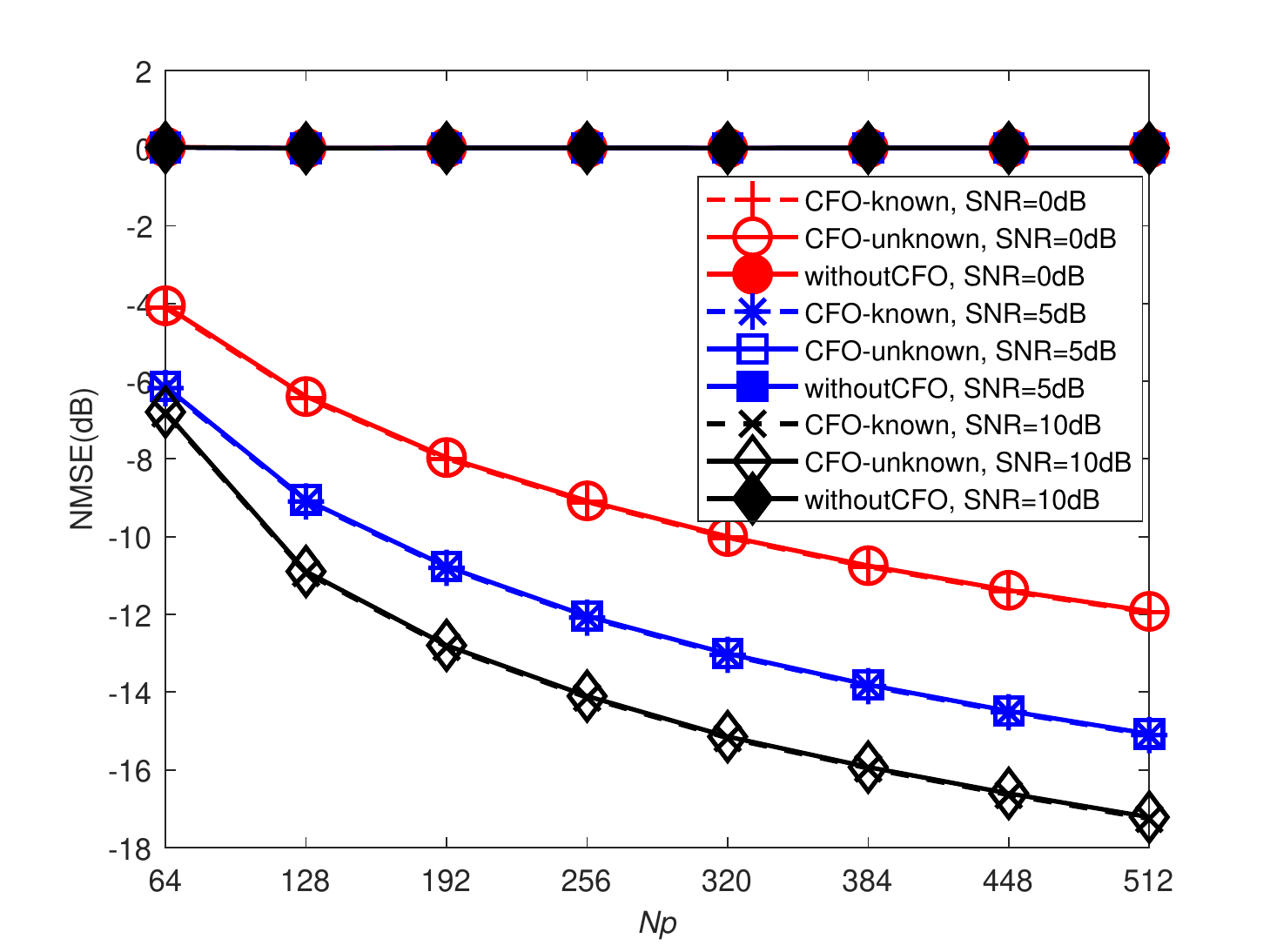}}
\caption{The NMSE of the channel versus the length of training block $N_p$ with different SNRs for the general Gaussian channel.
}\label{MSEh_Gaussian}
\end{figure}

In this experiment, the MSE of CFO and the NMSE of channel are compared with the corresponding CRB and CFO-known algorithm, respectively. The results are presented in Fig. \ref{MSEw_Gaussian} and Fig. \ref{MSEh_Gaussian}. In Fig. \ref{MSEw_Gaussian}, the MSE of CFO decreases as the length of training block $N_p$ increases or the SNR increases. And the performance gap between the MSE and CRB is less than about $3$dB. When the refinement step in the CFO estimation is not applied, the algorithm's performance is much worse. Fig. \ref{MSEh_Gaussian} presents the NMSE of channel with unknown CFO, known CFO and without the CFO compensation (assuming CFO is zero in the channel estimation) respectively. It can be seen that the NMSE of channel decreases when $N_p$ or SNR increases. And the NMSE of channel with unknown CFO is close to that with known CFO. Furthermore, when there is no CFO compensation, we can see that the channel can not be recovered successfully.

\subsection{mmWave Channel}
\begin{figure}
\centering{\includegraphics[width=2.5in]{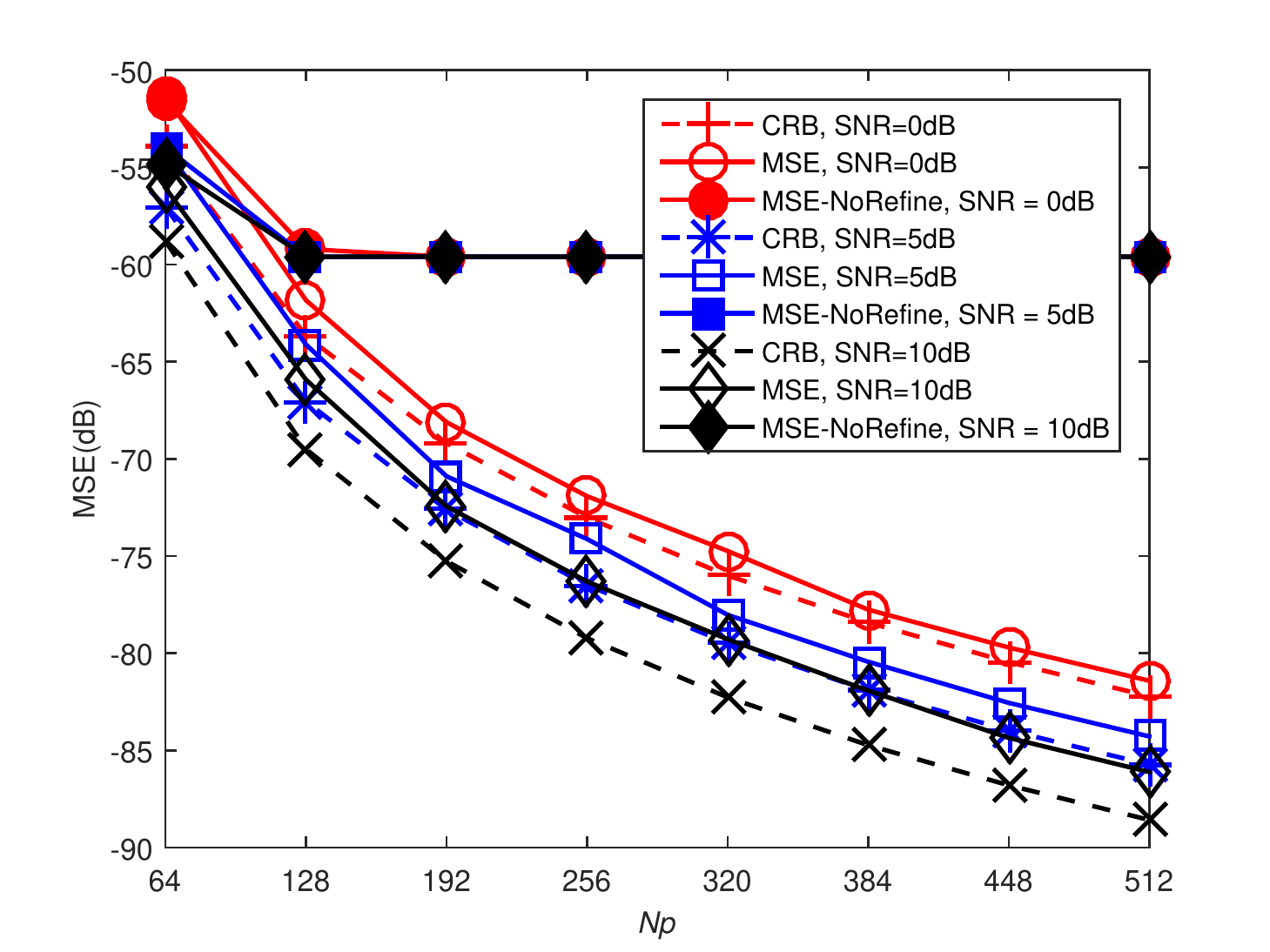}}
\caption{The MSE and CRB of the CFO versus the length of training block $N_p$ with different SNRs for the mmWave channel.
}\label{MSEw_mmWave}
\end{figure}

\begin{figure}
\centering{\includegraphics[width=2.5in]{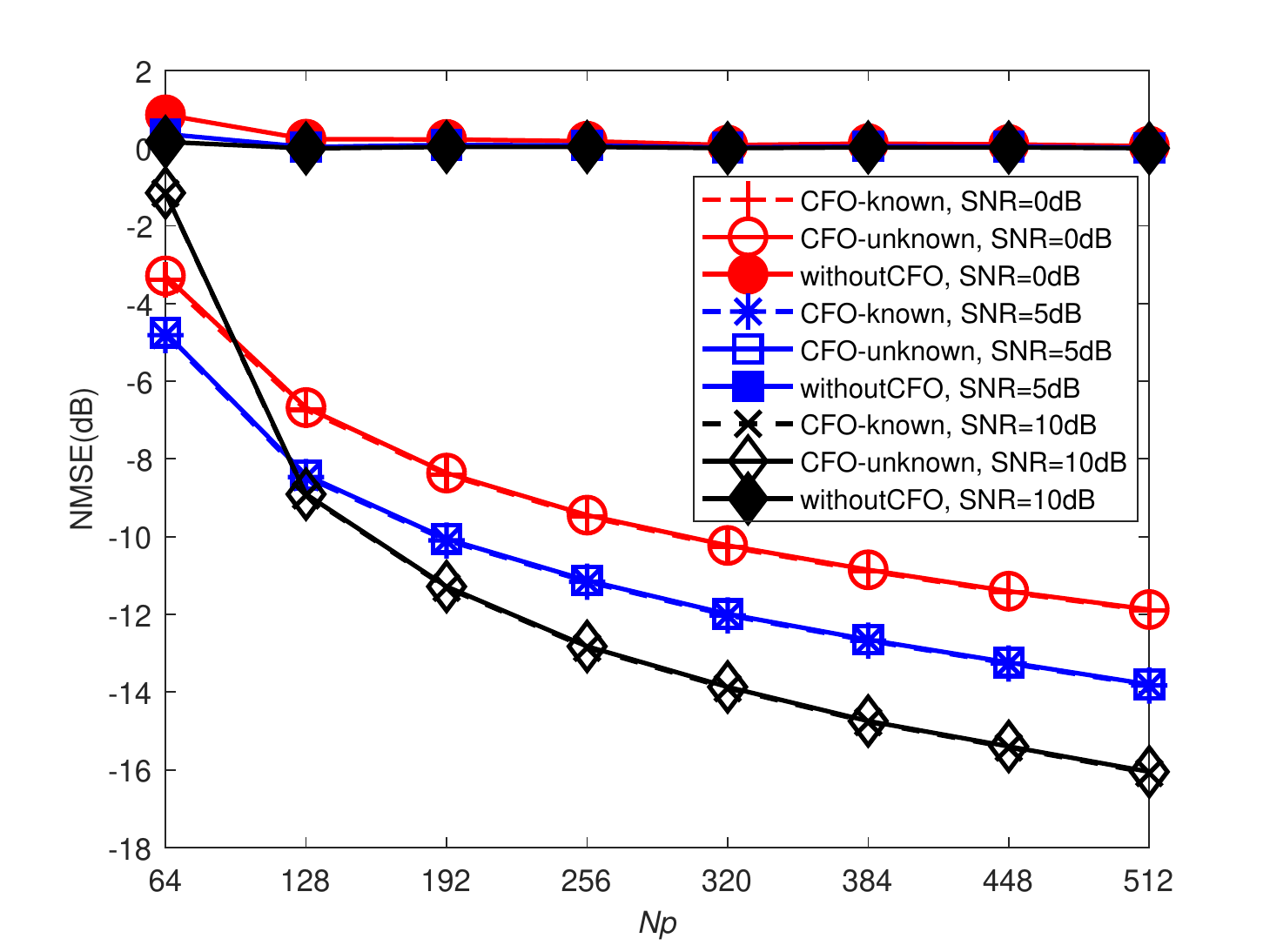}}
\caption{The NMSE of the channel versus the length of training block $N_p$ with different SNRs for the mmWave channel.
}\label{MSEh_mmWave}
\end{figure}

For the mmWave channel, we set the parameters of channel as follows: $N_c=2$, $K_n=15$ and $d = \lambda/2$. We generate $\theta_{r,m,n}$ and $\theta_{t.m.n}$ for the Laplacian distribution with an angle spread of 10 degrees \cite{Myers}. Numerical results are presented in Fig. \ref{MSEw_mmWave} and Fig. \ref{MSEh_mmWave}.

From Fig. \ref{MSEw_mmWave} and Fig. \ref{MSEh_mmWave}, we can see that the performance of our proposed approach for mmWave channel is similar to that for general Gaussian channel, which demonstrates that the proposed approach is effective for the mmWave channel.

For the mmWave channel, we also make a performance comparison with the approach in \cite{Myers} but only for the $N_p=64$ case. Because for the approach in \cite{Myers}, it's not practical to run it for a larger $N_p$. When $N_p=64$ and ${\rm SNR}=10{\rm dB}$, the MSE of the CFO of the approach in \cite{Myers} is about $-31{\rm dB}$ while our algorithm achieves $-55{\rm dB}$. The poor performance of the approach in \cite{Myers} can be attributed to the use of a discrete grid for CFO in the lifted vector. Increasing the resolution of the DFT grid in \cite{Myers} for a better CFO estimate, however, significantly increases the complexity of the algorithm in \cite{Myers}.

Besides, the running time of the proposed approach in both the general Gaussian channel and mmWave channel is shown in TABLE \ref{RunTime}. All results are obtained by an ordinary PC with an Intel Core i7 3.40 GHz CPU and 64.0 GB RAM. The value in parenthesis represents the running time of the algorithm for mmWave channel. Simulations with the CFO changing is referred to the supplementary materials.
\begin{table}[h!t]
    \begin{center}
\caption{The averaged running time (seconds) of the proposed approach.}\label{RunTime}
    \begin{tabular}{|c|c|c|c|c|c|c|c|c|}
            \hline
            Estimation$\backslash$ $N_p$& $64$ & $256$ & $512$\\ \hline
            CFO est. &1.1(1.1)&4.5(4.5)&9.8(9.8)\\ \hline
            Channel est.&0.1(0.3)&0.3(1.2)&0.6(2.2)\\ \hline
            Total&1.2(1.3)&4.8(5.6)&10.4(12.0)\\ \hline
        \end{tabular}
    \end{center}
\end{table}
%
\section{Conclusion}
We have designed a two-stage approach to estimate CFO and channel with one-bit ADCs, and derived the CRB of CFO. Numerical results demonstrate that the proposed approach works well for both the general Gaussian channel and mmWave channel, and the gap between the MSE and CRB of CFO is less than about 3dB. Compared to the CFO-known GAMP-EM algorithm, the performance degradation of the proposed approach is negligible.
\section{Acknowledgement}
We thank all the reviewers for their valuable comments which helps us improve this work. Also, we thank Nitin Jonathan Myers for sharing his codes and discussing with us, which helps us complete the comparison.
\section{Supplementary Material}
\subsection{Parameters $N_1$ and $N_2$}
In this subsection, the way to empirically choose the parameters $N_1$ (the number of grids in coarse detection) and $N_2$ (the number of grids in refined detection) is presented. For the convenience, the parameters of the numerical experiments in this material are set the same as that in our letter for the general Gaussian channel and SNR is chosen to be 10dB.
\begin{figure*}[!ht]
\centering
\subfigure[$N_p=64$]{
\label{Np64} 
\includegraphics[width=2.1in,height=1.8in]{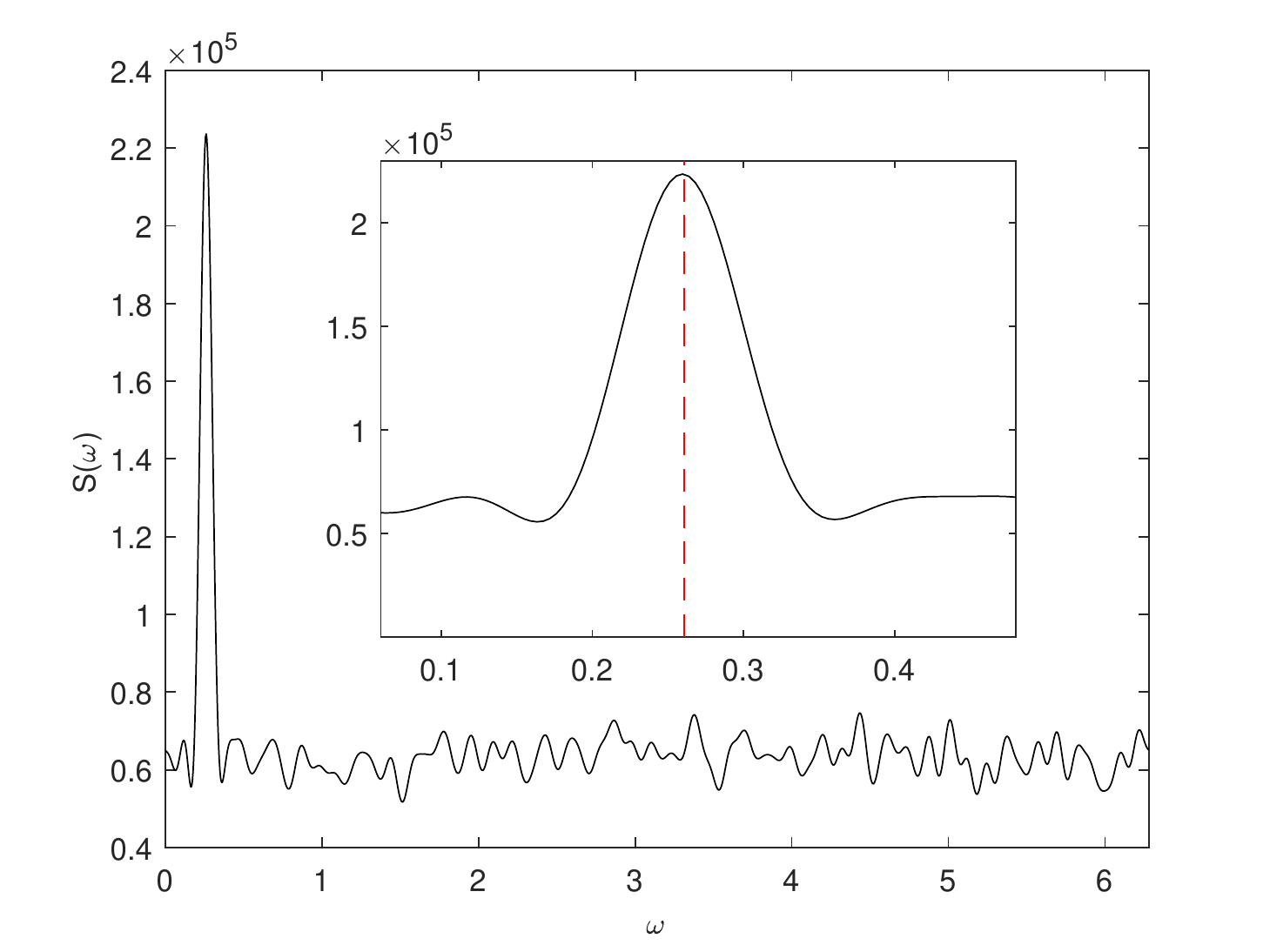}}
\hspace{0.1in}
\subfigure[$N_p=256$]{
\label{Np256} 
\includegraphics[width=2.1in,height=1.8in]{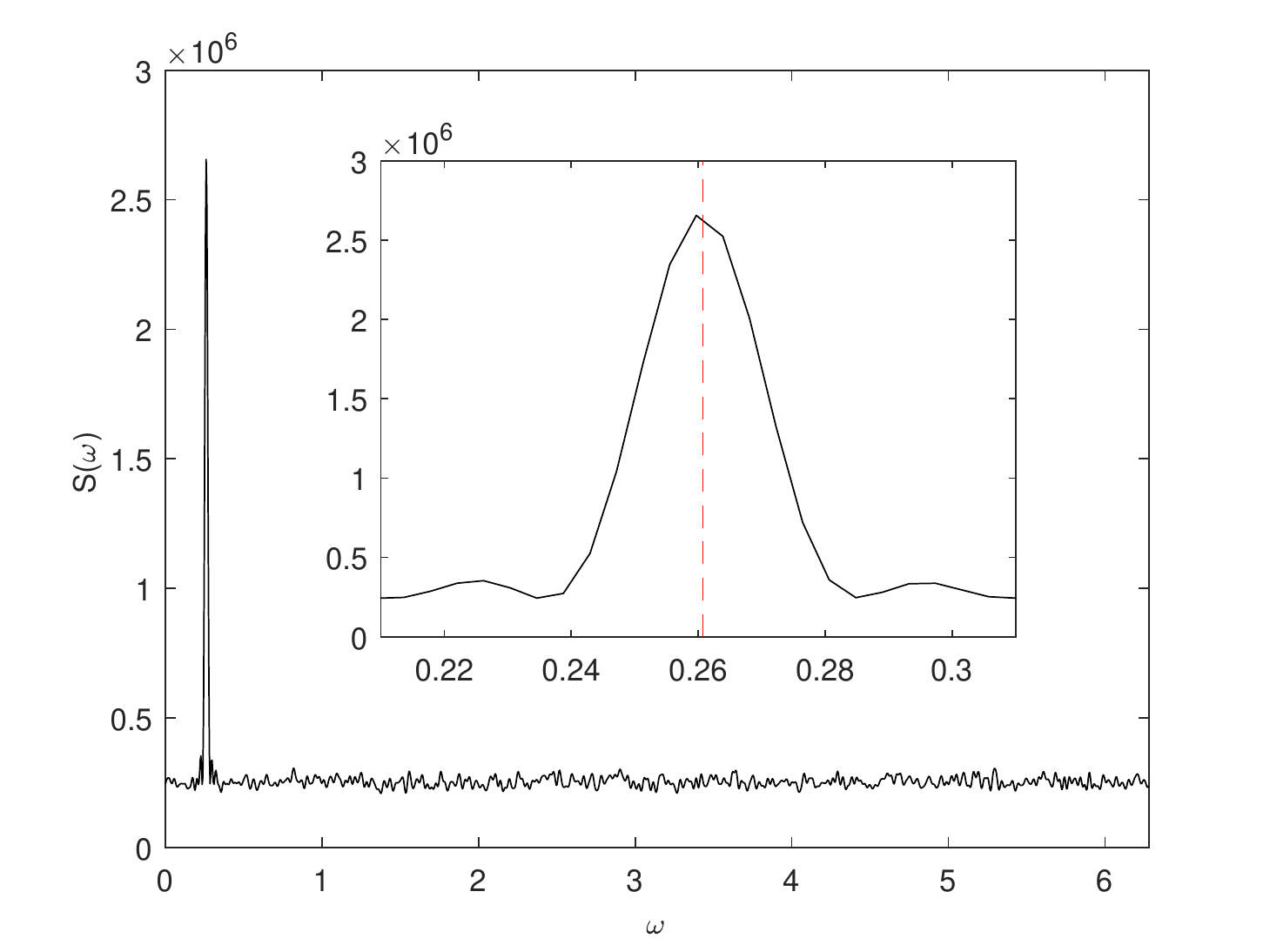}}
\hspace{0.1in}
\subfigure[$N_p=512$]{
\label{Np512} 
\includegraphics[width=2.1in,height=1.8in]{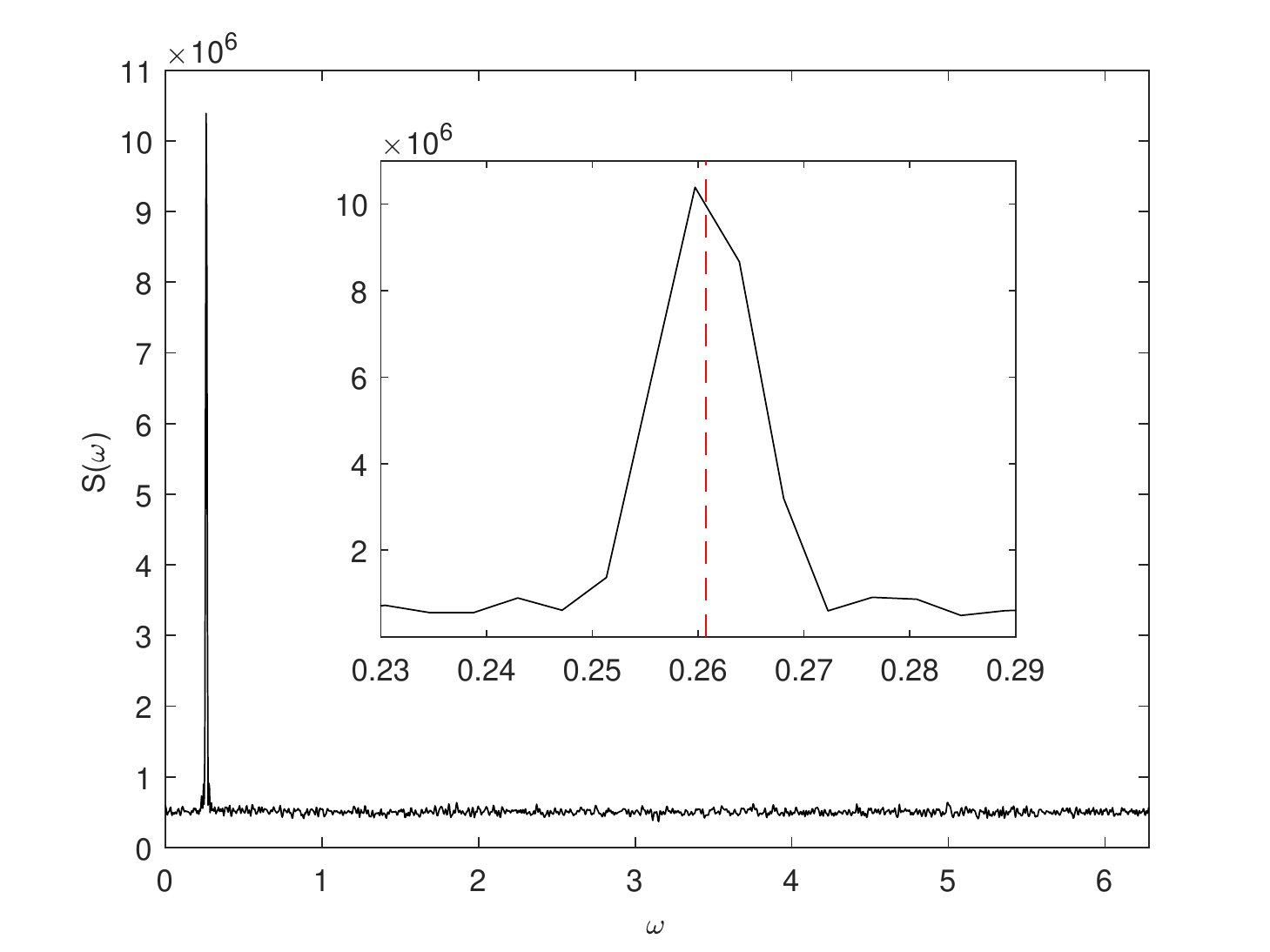}}
\centering
\caption{The curve of the objective function $S(\omega)$ with different $N_p$. The red dash line represents the true CFO $\omega_e$.}
\label{generalfigure} 
\end{figure*}
Let $S(\omega)={\Vert{\mathbf D}^{\rm T}\mathbf y\Vert}_2^2$, Fig. \ref{generalfigure} plots the curves of $S(\omega)$ with different $N_p$. From Fig. \ref{generalfigure}, we can see that there is an obvious main lobe and it is locally concave around the global optimum. Therefore, as long as $N_1$ is large enough, we can detect a value of $\omega$ on the main lobe which can make the gradient descent algorithm converge to the global optimum. However, in order to reduce the computational complexity, we choose $N_1$ reasonably which works well for the Detection step.

\begin{figure}[h!t]
\centering
{\includegraphics[width=2.8in]{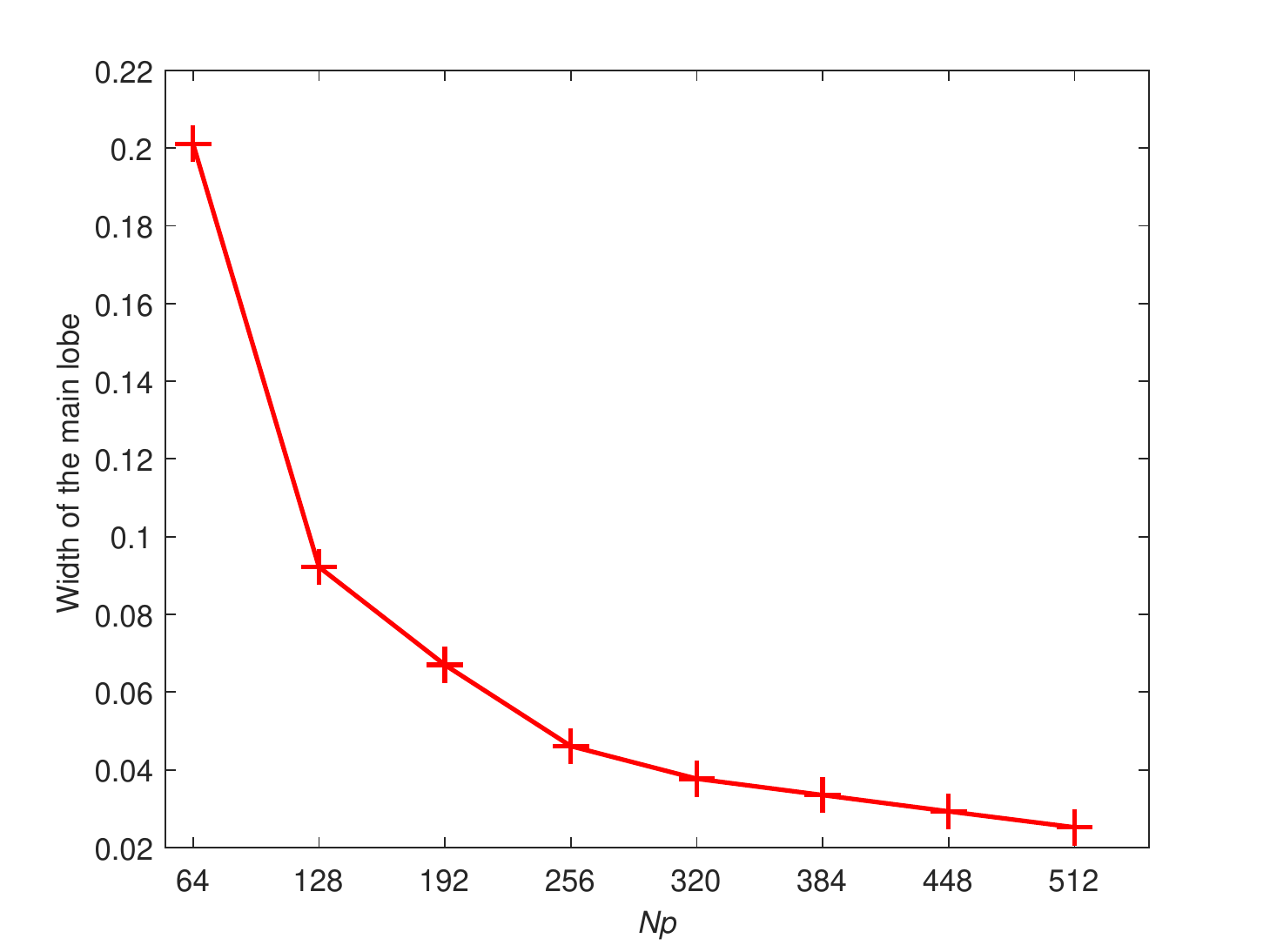}}
\caption{The width of the main lobe versus the length of training block $N_p$.
}\label{lobe}
\end{figure}

Through a large number of experiments, we found that the width of the main lobe depends mainly on $N_p$. Fig. \ref{lobe} presents the width of the main lobe with different $N_p$. It can be seen that the width of the main lobe decreases when $N_p$ increases and the smallest width of the main lobe is about 0.025. Thus, $N_1$ should be at least larger than $\frac{2\pi}{0.025}\approx 251$. However, we can see that there are many side lobes close to the main lobe, and meanwhile the main lobe in figures may also contain invisible side lobes, which may degrade the detection performance. Therefore, we design a two-step Detection: coarse detection and refined detection, and choose $N_1=300$ and $N_2=10$ for the simulations.



\subsection{Additional Numerical Simulation}
\begin{figure}[h!t]
\centering{\includegraphics[width=2.8in]{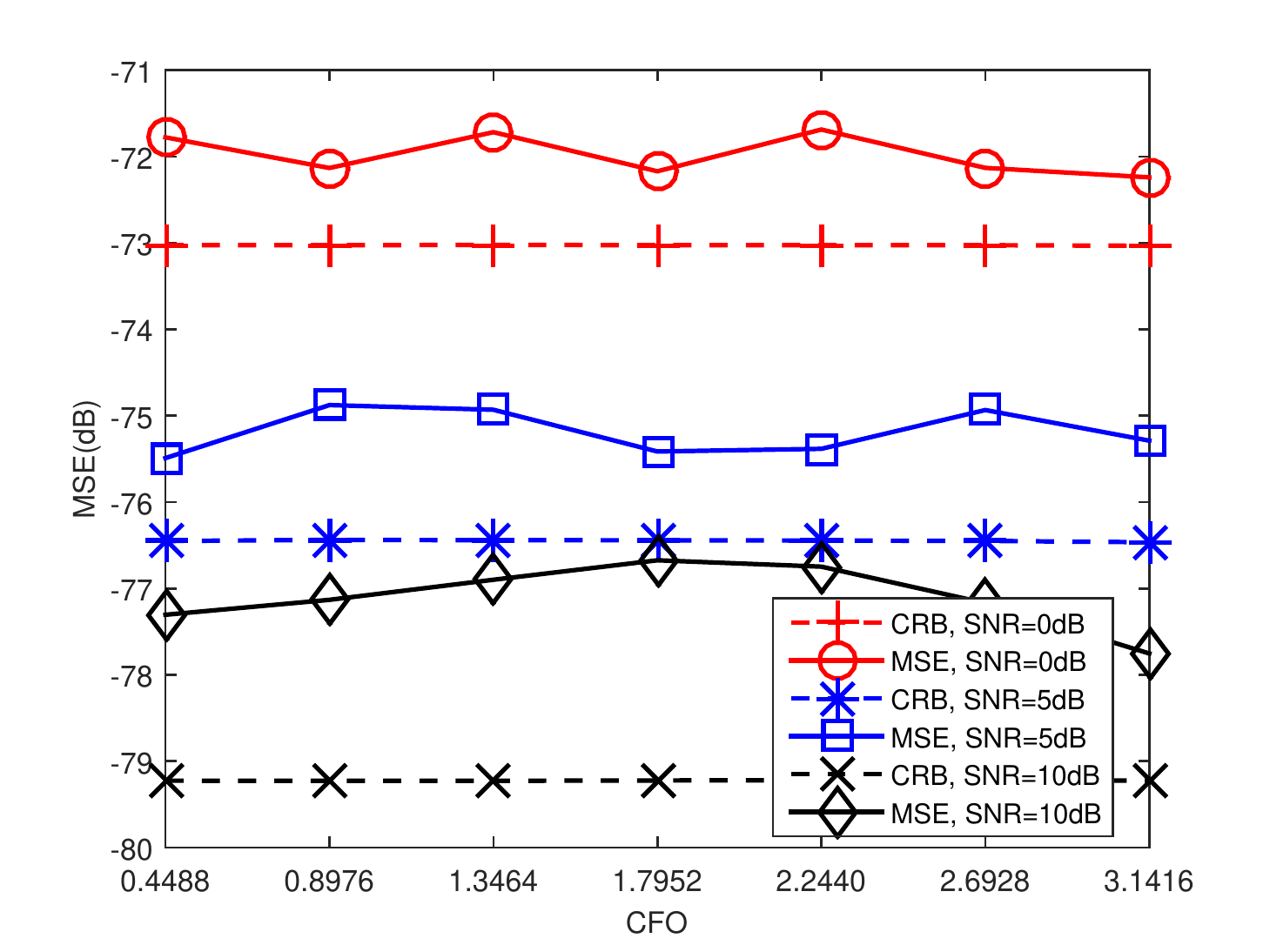}}
\caption{The MSE of the CFO versus the CFO $\omega_e$ with $N_p=256$ and ${\rm SNR}=10{\rm dB}$ for the general Gaussian channel.
}\label{MSEw_gc_CFOchange}
\end{figure}

\begin{figure}[h!t]
\centering{\includegraphics[width=2.8in]{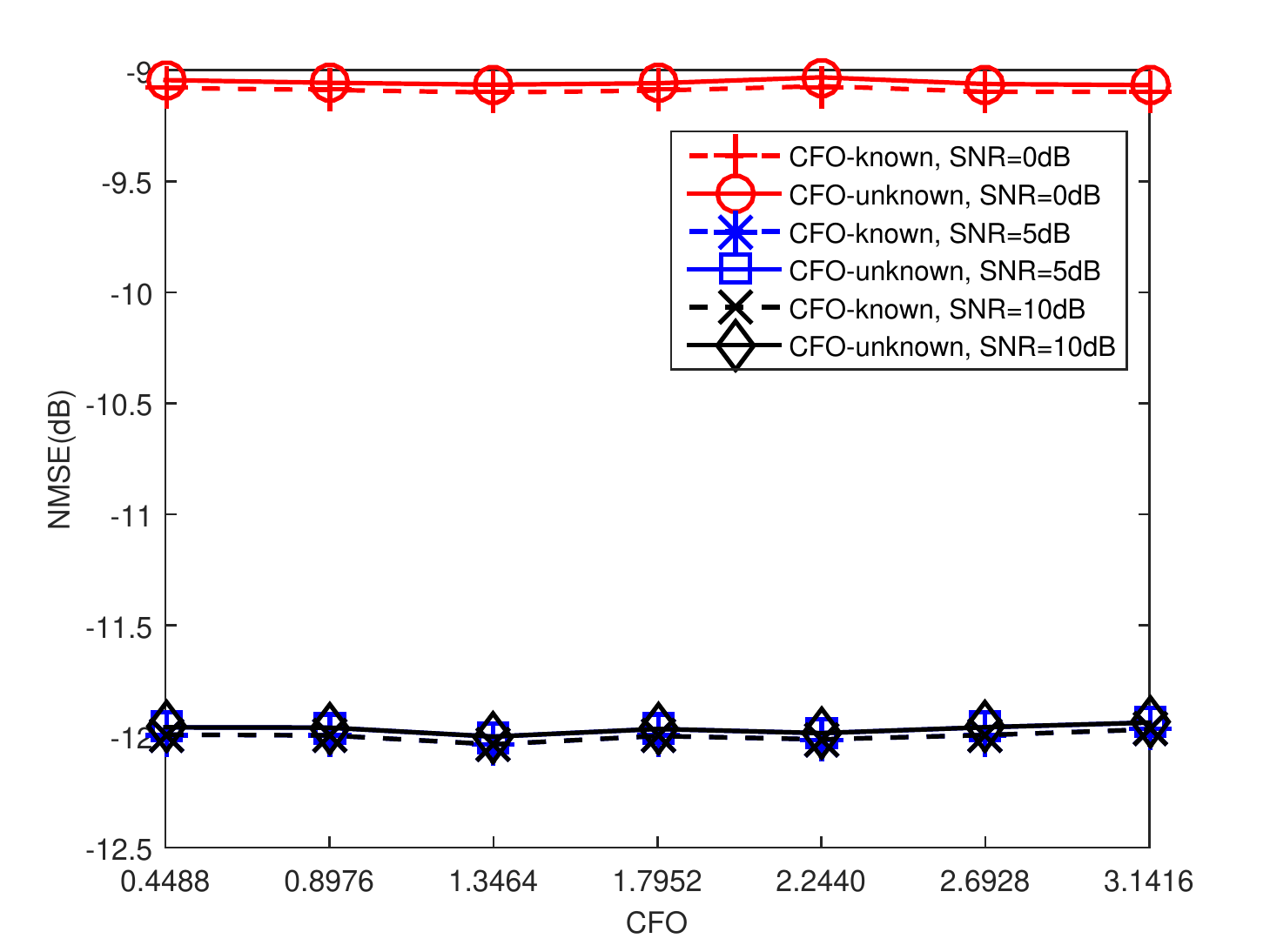}}
\caption{The NMSE of the channel versus the CFO $\omega_e$ with $N_p=256$ and ${\rm SNR}=10{\rm dB}$ for the general Gaussian channel.
}\label{NMSEh_gc_CFOchange}
\end{figure}

\begin{figure}[h!t]
\centering{\includegraphics[width=2.8in]{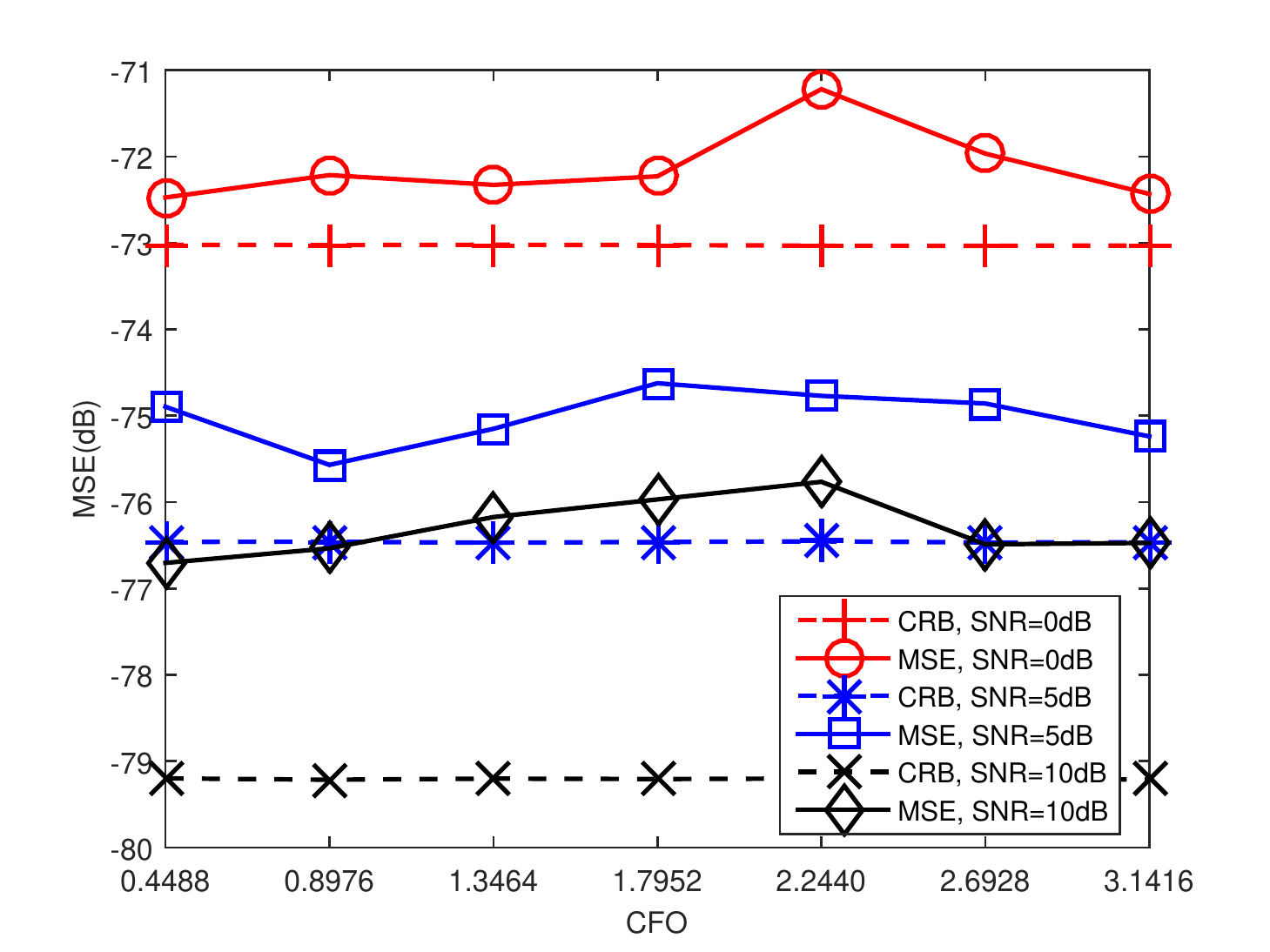}}
\caption{The MSE of the CFO versus the CFO $\omega_e$ with $N_p=256$ and ${\rm SNR}=10{\rm dB}$ for the general mmWave channel.
}\label{MSEw_mc_CFOchange}
\end{figure}

\begin{figure}[H]
\centering{\includegraphics[width=2.8in]{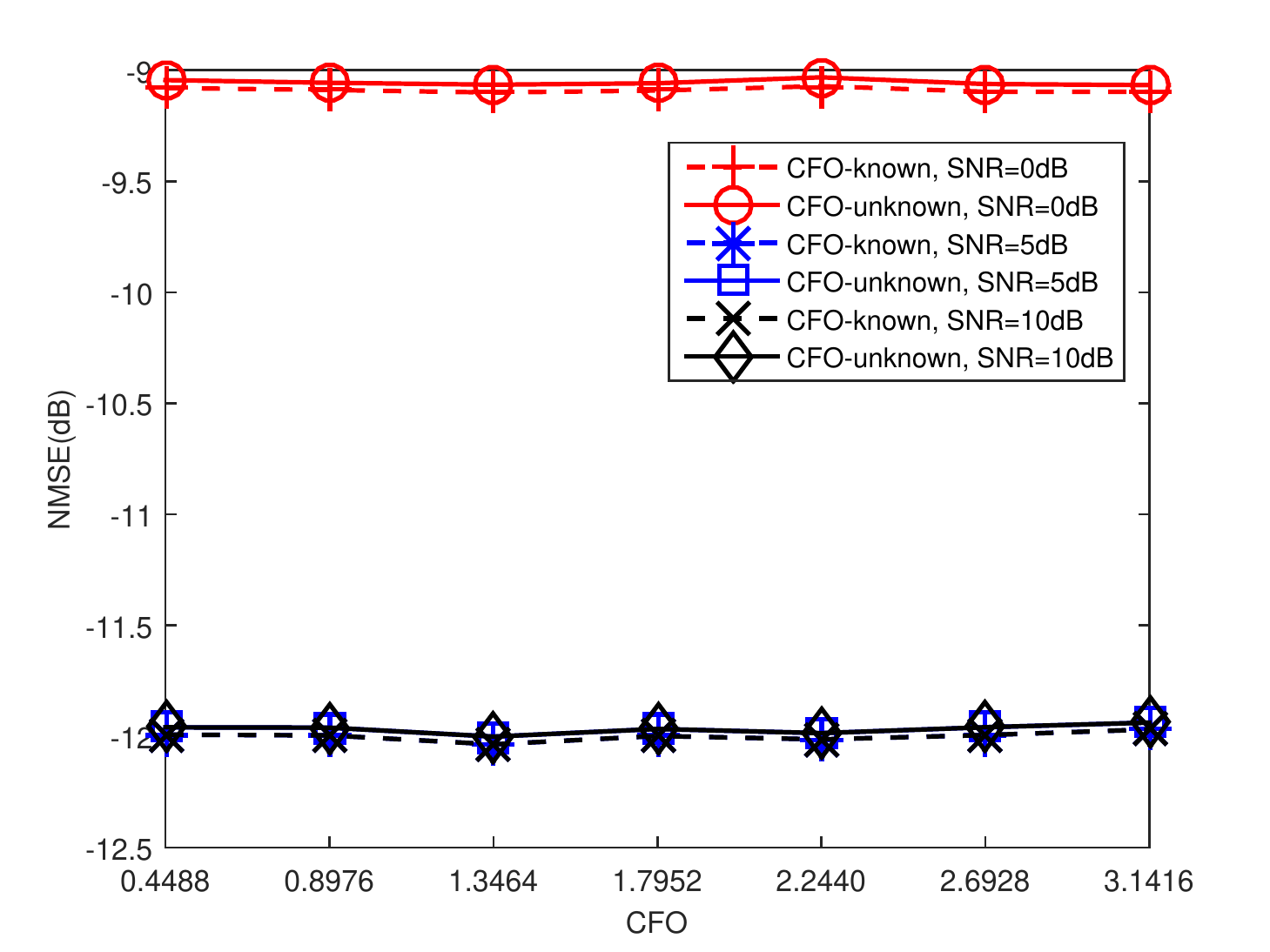}}
\caption{The NMSE of the channel versus the CFO $\omega_e$ with $N_p=256$ and ${\rm SNR}=10{\rm dB}$ for the general mmWave channel.
}\label{NMSEh_mc_CFOchange}
\end{figure}

The performance of the proposed algorithm with different values of CFO and fixed values of $N_p$ and SNR is provided. We choose $N_p=256$ and ${\rm SNR}=10{\rm dB}$ and all other parameters are set the same as that in the paper. The results are in Fig. \ref{MSEw_gc_CFOchange}, \ref{NMSEh_gc_CFOchange}, \ref{MSEw_mc_CFOchange} and \ref{NMSEh_mc_CFOchange}. It can be seen that the proposed algorithm performs well for all values of CFO for both the Gaussian channel and the mmWave channel. Besides, with the CFO changing, both the MSE of the CFO estimation and the channel estimation are stable.

\end{document}